\documentstyle[12pt]{article}

\begin{document}
\thispagestyle{empty}

\begin{center}
               RUSSIAN GRAVITATIONAL SOCIETY\\
               INSTITUTE OF METROLOGICAL SERVICE \\
               CENTER FOR GRAVITATION AND FUNDAMENTAL METROLOGY\\

\end{center}
\vskip 4ex
\begin{flushright}
                                         RGS-VNIIMS-007/95
                                         \\ gr-qc/9510059

\end{flushright}
\vskip 15mm

\begin{center}
{\large\bf MULTIDIMENSIONAL GENERALIZATION OF KASNER SOLUTION }

\vskip 5mm
{\bf
Sergey S. Kokarev }\\
\vskip 5mm
{\em 150000, Yaroslavl, Respublikanskaya 108, Phys.-Math. Department,r.409} \\
e-mail: glad@delta.yaroslavl.su    \\
\end{center}
\vskip 10mm

\begin{abstract}
Full generalization of Kasner metric for the case of $n+1$ dimensions and
$m\le n+1$ essential variables is obtained. Any solution is defined by the
corresponding constant matrix of Kasner parameters. This parameters
form in euclidian space Casner hyperspheres and are connected by additional
conditions. General properties of obtained solutions are analised.
\end{abstract}

\vfill

\centerline{Moscow 1995}
\pagebreak

\section{General solution}.

In present paper we'll find full multidimensional generalization of Kasner
solution, which in common representation has the following kind:
\begin{equation}\label{kaz4}
ds^{2}=dt^{2}-t^{2p_{1}}dx^{2}-t^{2p_{2}}dy^{2}-t^{2p_{3}}dz^{2},
\end{equation}
which describe homogeneous anisotropic 4-D vacuum space-time.
Here $p_{1},p_{2},p_{3}$ --- real parameters, satisfying to the following
conditions:
\begin{equation}\label{cd}
p_{1}+p_{2}+p_{3}=p_{1}^{2}+p_{2}^{2}+p_{3}^{2}=1.
\end{equation}
Original form of Kasner solution, given in \cite{kaz}, is more general and
has the following kind:
\begin{equation}
ds^{2}=t^{2p_{0}}dt^{2}-t^{2p_{1}}dx^{2}-t^{2p_{2}}dy^{2}-t^{2p_{3}}dz^{2}
\end{equation}
with conditions on parametres $p_{1}+p_{2}+p_{3}=1+p_{0}$, \\
$p_{1}^{2}+p_{2}^{2}+p_{3}^{2}=(1+p_{0})^{2}$
In general, there is freedom in choosing of value of parameter $p_{0}$ by
the coordinate transformations.  Another commonly used form of this solution
have been proposed by Narlikar and Karmarkar \cite{nar}.
4-D Casner solution is suitable for describing of early epoch of the Universe
\cite{ll}.

Multydimensional generalization of metric (\ref{kaz4}) for the case of $n+1$-D
homogeneous anisotropic space-time has been found in  \cite{iv}.
It has the following kind:
\begin{equation}\label{kazmd}
ds^{2}=dt^{2}-\sum\limits_{k=1}^{n}t^{2p_{k}}(dx^{k})^{2},
\end{equation}
with the conditions on parameters $p_{k}$:
\begin{equation}
\sum\limits_{k=1}^{n}p_{k}=\sum\limits_{k=1}^{n}p_{k}^{2}=1.
\end{equation}

We'll find generalization of the metric (\ref{kazmd}), when it depends on
degrees of another coordinates (all or some parts) with their own Casner
parameters.
Such space will not be homogeneous now, but it will admits the system of
$n+1$-orthogonal hyphersurface \cite{eiz1},
which in choosen coordinate system will be coordinate hyphersurfaces
$x^{i}=const$.

Multidimensional metric we can write in the following symmetric form:
\begin{equation}
ds^{2}=e^{2\alpha_{0}}dt^{2}-\sum\limits_{k=1}^{n}e^{2\alpha_{k}}(dx^{k})^{2},
\end{equation}
where $\{\alpha_{0},\dots\alpha_{n}$--- are, in general, functions of
(without loss of generality) first $m$ coordinates
$\{t,x^{1},\dots,x^{m}\}$. Multydimensional vacuum Einstein equations\\
${}^{n+1}\!R_{ij}=0$ have the following kind:
\begin{equation}\label{eqkazdg}
R_{ii}=\sum\limits_{k}'\varepsilon_{k}e^{-2\alpha_{k}}
\left(\alpha_{i,kk}+\alpha_{i,k}\times\right.
\end{equation}
\[
\left.\left(\sum'\limits_{p}\alpha_{p,k}
-\alpha_{k,k}\right)\right)+\varepsilon_{i}e^{-2\alpha_{i}}\left(\sum'\limits_{s}
\alpha_{s,ii}+\right.
\]
\[
\left.\left.\alpha_{s,i}(\alpha_{s,i}-\alpha_{i,i}\right)\right)=0;
\]
\[
i=\overline{0,m},\ k=\overline{0,n}\ (k\neq i),\ p=\overline{0,n}\ (p\neq k),
\]
\[
s=\overline{0,n}\ (s\neq i).
\]

\begin{equation}\label{eqkazndg}
R_{ij}=e^{-\alpha_{i}-\alpha_{j}}\left(-\sum''\limits_{k}\alpha_{k,ij}+
\alpha_{i,j}\sum''\limits_{k}\alpha_{k,i}+\right.
\end{equation}
\[
\left.\alpha_{j,i}\sum''\limits_{k}\alpha_{k,j}-
\sum''\limits_{k}\alpha_{k,i}\alpha_{k,j}\right)=0.
\]
\[
k=\overline{0,n}\ (k\neq i,j)
\]
where symbols "'" and "''" means, that summating is made by all set of
corresponding
index, excluding one or two values of this set correspondingly.
We are using the following abbreviations:
\[
\alpha_{i,j}\equiv\frac{\partial\alpha_{i}}{\partial x^{j}};\ \
\alpha_{i,jk}\equiv\frac{\partial^{2}\alpha_{i}}{\partial x^{j}\partial x^{k}}.
\]
\[
\varepsilon_{k}=\left\{\begin{array}{ll}
+1,&\ \ \ k=0;\\
-1,&\ \ \ k\neq0.
\end{array}
\right.
\]
Obviously, that all partial derivatives $\partial/\partial_{k}$ with $k>m$
vanishes
in eq. (\ref{eqkazdg})-(\ref{eqkazndg}).
For we are interested by generalization of the solution
(\ref{kazmd}), then it is necessary to specify kind of the functions
$\alpha_{k}$ by the following way:
\begin{equation}\label{matr}
\alpha_{k}=\alpha_{ks}\ln x_{s}
\end{equation}
where $\alpha_{ks}$ --- is a $n\times m$ constant matrix of Casner parameters.
In (\ref{matr}) there is summating by repeating index $s=\overline{0,m}$.
Substituting this kind of functions  $\alpha_{k}$ into
system (\ref{eqkazdg})-(\ref{eqkazndg}) and taking into account
that  $\alpha_{k,i}=\alpha_{ki}/x^{i},\ \ \
\alpha_{k,ii}=-\alpha_{ki}/(x^{i})^{2},\ \ \ \alpha_{k,ij=0}$ we  can get
the following system of equation:

\begin{equation}\label{geqd}
\sum'\limits_{k}
\frac{\varepsilon_{k}\alpha_{ik}}{(x^{k})^{2}e^{2\alpha_{k}}}
\left(-1+\sum'\limits_{p}\alpha_{pk}-\alpha_{kk}\right)+
\end{equation}
\[
\frac{\varepsilon_{i}}{(x^{i})^{2}e^{2\alpha_{i}}}
\left(-\sum'\limits_{s=0}\alpha_{si}+
\alpha_{si}(\alpha_{si}-\alpha_{ii})\right)=0;
\]

\begin{equation}\label{geqnd}
\frac{e^{-\alpha_{i}-\alpha_{j}}}{x^{i}x^{j}}\left(\alpha_{ij}\sum''\limits_{k}\alpha_{ki}+
\alpha_{ji}\sum''\limits_{k}\alpha_{k,j}-\sum''\limits_{k}\alpha_{ki}\alpha_{kj}\right)
\end{equation}
\[
=0.
\]
Diagonal equations (\ref{geqd})
gives the restrictions on Casner parameters
for each variable, that can be put to the form:
\begin{equation}\label{cond1}
\sum\limits_{p=0}^{n}\alpha_{pk}=1+2\alpha_{kk};\
\sum\limits_{p=0}^{n}(\alpha_{pk})^{2}=
1+2\alpha_{kk}+2\alpha_{kk}^{2};
\end{equation}
\[
(k=\overline{0,m})
\]
Nondiagonal equations (\ref{geqnd}) give supplement cross restrictions on
different pair of Casner parameters:
\begin{equation}\label{cond2}
\!\!\sum\limits_{p=0}^{n}\alpha_{pi}\alpha_{pj}=
\alpha_{ij}(2\alpha_{ii}+1)+\alpha_{ji}(2\alpha_{jj}+1)-2\alpha_{ij}\alpha_{ji};
\end{equation}
\[
(i,j=\overline{0,m},\ i\not=j)
\]

By the suitable coordinate
transformation parameters $\alpha_{ii}$ can be transformed into any real value.
Paticulary, if $\alpha_{ii}=0$, we come to the common conditions as in
(\ref{kazmd})  parameters $\alpha_{ij}$. To simplify conditions
(\ref{cond2}) we take $\alpha_{ii}=-1/2$. Then
(\ref{cond1})---(\ref{cond2}) take the following form:
\begin{equation}\label{cond}
\sum\limits_{p=0}^{n}\alpha_{pk}=0;\  \ \
\sum\limits_{p=0}^{n}\alpha_{pk}^{2}=\frac{1}{2};
\end{equation}
\begin{equation}\label{crcd}
\sum\limits_{p=0}^{n}\alpha_{pi}\alpha_{pj}=-2\alpha_{ij}\alpha_{ji}.
\end{equation}
So, our generalised solution has the following form:
\begin{equation}\label{sol}
ds^{2}=\prod\limits_{s=0}^{m}(x^{s})^{2\alpha_{0s}}dt^{2}-\sum\limits_{k=1}^{n}
\prod\limits_{s=0}^{m}(x^{s})^{2\alpha_{ks}}(dx^{k})^{2}
\end{equation}
with  conditions (\ref{cond})-(\ref{crcd}).

\section{Dimension of manifold of Kasner parameters}.

It is easy to note firstly, that any generalized Kasner solution (\ref{sol})
characterized by the its own matrix of parameters:
\begin{equation}
\left(
\begin{array}{llll}
-1/2        & \alpha_{02}  & \cdots & \alpha_{0m}\\
\alpha_{10} & -1/2         & \cdots & \alpha_{1m}\\
\vdots      & \vdots       & \vdots & \vdots     \\
\alpha_{n0} & \alpha_{n1}  & \cdots & \alpha_{nm}
\end{array}
\right),
\end{equation}
where number of lines is equal to $n+1$ --- dimension of space-time,
and number of coloumns is equal to the $m$ --- number of
variables, which multydimensional metric depend on.
Secondly, to every solution some surface in euclidian $mn$-dimensional
space of parameters $\{\alpha_{ks}\}$ corresponds.
This surface is determined by the equations (\ref{cond})-(\ref{crcd}).
Its dimension can be easily founded by taking difference
between number of parameters --- $mn$ and the number of eq.
(\ref{cond})-(\ref{crcd})
--- $m(m+3)/2$. Result $m(2n-m-3)/2$ can be written in the form of
nonequality, which is a sequence of demanding, that
dimension of the surface will be more or equal to zero
(or system (\ref{cond})-(\ref{crcd}) will be simultaneous):
\begin{equation}
n\ge\frac{m+3}{2}.
\end{equation}
Then we at once get the following infinite table,
showing relation between dimension of space-time and number
of variables in metric with dimension of parametric surface
of parameters subspace:
\begin{center}
\begin{tabular}{|c|c|c|c|c|c|c}
\hline
$n\backslash m$ & 1 & 2 & 3 & 4 & 5 & $\cdots$ \\ \hline
0               & - &   &   &   &   & $\cdots$ \\ \hline
1               & - & - &   &   &   & $\cdots$ \\ \hline
2               & 0 & - & - &   &   & $\cdots$ \\ \hline
3               & 1 & 1 & 0 & - &   & $\cdots$ \\ \hline
4               & 2 & 3 & 3 & 2 & 0 & $\cdots$ \\ \hline
5               & 3 & 5 & 6 & 6 & 5 & $\cdots$ \\ \hline
$\vdots$          &   &   &$\cdots$&   &   & $\cdots$
\end{tabular}
\end{center}
Here in left coloumn put the dimension $n$ of  $n+1$ -dimensional
space-time is laid off, in top horizontal line --- the number of coordinates in
metric,
in table --- corresponding dimension of subspace of parameters.
Region of the table with empty cells corresponds to the obvious fact,that
number of variables, which metric depend  on, can not be more then dimension
of space-time.
Symbol "-" means,
that for given values of $n$ and $m$ dimension of subspace of parameters
is negative, i.e. there is no solution of eq.(\ref{cond})-(\ref{crcd}).
For the cases $n=0,1,2$ there is no solutions, because
for  $n<3$ there is no exist nontrivial vacuum solutions of Einstein equations.
For $n=3$, as it can been seen from the table, there is no solutions with
$m=4$.
As will be shown in sec.4 case with $m=3$ reduces to original Kasner solution
(\ref{kaz4}) with special values of parameters.
For another cases solutions with arbitrary  $m$ and $n$
for $m\le n+1$ are exist.

\section{Parametrization of Kasner hyphersphere}.

Let us consider two first equations (\ref{cond}).
For every variable $x^{k}$  this
two equations formally determine hyphersphere $S_{n-1}$ in $n+1$-dimensional
euclidian space of parameters (we remember that in our
parametrization $\alpha_{ii}=-1/2$, so we have in fact hyphersphere $S_{n-2}$
in
$n+1$-dimensional euclidian space of parameters). Take for example $k=0$ and
denote $\alpha_{k0}=p_{k}$ --- the components of radius-vector of
Kasner parameters.
So we can describe our hyphersphere $S_{n-1}$  by one vector equation
\begin{equation}\label{vec}
\vec p=p_{0}\vec e_{0}+p_{1}\vec e_{1}+\dots+p_{n}\vec e_{n}
\end{equation}
with equations on coordinates $p_{i}$:
\begin{equation}\label{hyp}
\left\{
\begin{array}{rcl}
p_{0}+p_{1}+\dots+p_{n}&=&0;\\
p_{0}^{2}+p_{1}^{2}+\dots+p_{n}^{2}&=&1/2.
\end{array}
\right.
\end{equation}
 From equations (\ref{hyp}) one can see, that radius-vector describes
intersection of sphere $S_{n}$ with radius $1/\sqrt{2}$ and plane with
normal vector oriented along bissectriss of $n+1$-dimensional
coordinate angle. This intersection will be hyphersphere $S_{n-1}$,
which we'll call {\it Kasner hyphersphere.}

Its parametric description can be obtained by the following way:
\begin{equation}\label{par}
\vec p=\overline{O}(n+1)\vec p'
\end{equation}
where $\vec p'$ --- radius-vector with euclidian norm $|\vec p'|=1/\sqrt{2}$,
lied in plane $p_{n}=0$, which can be written in the following form:
\begin{equation}
\vec p'=\frac{1}{\sqrt{2}}\vec n=\frac{1}{\sqrt{2}}\sum\limits_{k=0}^{n}
\vec e_{k}\cos\theta_{k-1}\times
\end{equation}
\[
\prod\limits_{s=k}^{n-1}\sin\theta_{s}|_{\theta_{n-1}=\pi/2}
\]
Here $\{\theta_{0},\dots,\theta_{n-1}\}$ --- set of spherical angles
of $n+1$-dimensional spherical coordinate system with unit radius.
By $\overline{O}(n+1)$ in (\ref{par}) denoted $(n+1)\times(n+1)$ orthogonal
matrix,
which transform unit normal vector of plane $p_{n}=0$ with coordinates
$(0,\dots,1)$ into unit normal vector of plane $p_{0}+\dots+p_{n}=0$
with coordinates $(1/\sqrt{n+1},\dots,1/\sqrt{n+1})$.
This matrix can be put to the following form:
\begin{equation}\label{matr1}
\left(
\begin{array}{cccccc}
s_{0},&0,&0,&\dots,&0,&\frac{1}{\sqrt{n+1}}\\
\\
a_{0},&s_{1},&0,&\dots,&0,&\frac{1}{\sqrt{n+1}}\\
\\
a_{0},&a_{1},&s_{2},&0,&\dots,&\frac{1}{\sqrt{n+1}}\\
\\
a_{0},&a_{1},&a_{2},&s_{3},&\dots,&\frac{1}{\sqrt{n+1}}\\
\\
\dots&\dots&\dots&\dots&\dots&\dots\\
\\
a_{0},&a_{1},&\dots&,a_{n-2},&s_{n-1},&\frac{1}{\sqrt{n+1}}\\
\\
a_{0},&a_{1},&\dots&,a_{n-2},&-s_{n-1},&\frac{1}{\sqrt{n+1}}
\end{array}
\right)
\end{equation}
where $a_{i}=-1/\sqrt{(n-i)(n-i+1)}$,\\
$s_{i}=\sqrt{(n-i)/(n-i+1)}.$
We give here matrixes for $n=3,4$:
\begin{equation}
\overline{O}(4)=\left(
\begin{array}{cccc}
\frac{\sqrt{3}}{2},&0,&0,&\frac{1}{2}\\
\\
-\frac{1}{2\sqrt{3}},&\sqrt{\frac{2}{3}},&0,&\frac{1}{2}\\
\\
-\frac{1}{2\sqrt{3}},&-\frac{1}{\sqrt{6}},&\frac{1}{\sqrt{2}},&\frac{1}{2}\\
\\
-\frac{1}{2\sqrt{3}},&-\frac{1}{\sqrt{6}},&-\frac{1}{\sqrt{2}},&\frac{1}{2}
\end{array}
\right);
\end{equation}

\[
\overline{O}(5)=
\]
\begin{equation}
\left(
\begin{array}{ccccc}
\frac{2}{\sqrt{5}},&0,&0,&0,&\frac{1}{\sqrt{5}}\\
\\
-\frac{1}{2\sqrt{5}},&\frac{\sqrt{3}}{2},&0,&0,&\frac{1}{\sqrt{5}}\\
\\
-\frac{1}{2\sqrt{5}},&-\frac{1}{2\sqrt{3}},&\sqrt{\frac{2}{3}},&0,&\frac{1}{\sqrt{5}}\\
\\
-\frac{1}{2\sqrt{5}},&-\frac{1}{2\sqrt{3}},&-\frac{1}{\sqrt{6}},&\frac{1}{\sqrt{2}},&\frac{1}{\sqrt{5}}\\
\\
-\frac{1}{2\sqrt{5}},&-\frac{1}{2\sqrt{3}},&-\frac{1}{\sqrt{6}},&-\frac{1}{\sqrt{2}},&\frac{1}{\sqrt{5}}
\end{array}
\right)
\end{equation}

For the case of $m$ variables which metric depend on we'll have
parametric space in the form of direct production of $m$  copies of Kasner
hypherspheres:
$\underbrace{S_{n-1}\otimes S_{n-1}\cdots\otimes S_{n-1}}_{m}$ with supplement
cross conditions, given by the equations (\ref{crcd}),
which connect pairs of parametres from different Kasner hypherspheres.

\section{4-dimensional generalization of Kasner solution.}

Consider in details 4-dimensional generalization of Kasner solutions.
In all above formulaes we should  put $n=3$. For the case
$m=1$ we go to the metric (\ref{kaz4}) with 1-dimensional
parametric space.
Using the formalism of parametrization given in the previous section
we can get the following radius-vector $\vec p$:
\begin{equation}\label{hyp4}
\vec{p}=\left(
\begin{array}{l}
-\frac{1}{2}\\
\\
\frac{1}{6}-\frac{\sqrt{2}}{3}\cot\theta_{0}\\
\\
\frac{1}{6}+\frac{1}{3\sqrt{2}}\cot{\theta_{0}}+\frac{1}{2}\sqrt{\frac{1}{3}-
\frac{2}{3}\cot^{2}\theta_{0}}\\
\\
\frac{1}{6}+\frac{1}{3\sqrt{2}}\cot{\theta_{0}}-\frac{1}{2}\sqrt{\frac{1}{3}-
\frac{2}{3}\cot^{2}\theta_{0}}
\end{array}
\right)
\end{equation}
where parameter $\theta_{0}$ satisfy to the following conditions:
$-1/\sqrt{2}\le\cot\theta_{0}\le1/\sqrt{2}$.

For the case $m=2$ we have  two Kasner hypherspheres. One of them can be
parametrized
by the same way as (\ref{hyp4}). Parametrization of another hyphersphere
obtained from (\ref{hyp4}) by the interchange of $p_{0}$ and $p_{1}$
and by replacement $\theta_{0}\to\theta_{1}$. Additional cross condition
give the following equations on parameters $\theta_{0}$ and $\theta_{1}$:
\begin{equation}\label{eq}
-1+2(\xi+\eta)+5\xi\eta+3\sqrt{1-\xi^{2}}\sqrt{1-\eta^{2}}=0.
\end{equation}
where $\xi=\sqrt{2}\cot\theta_{0},\ \ \eta=\sqrt{2}\cot\theta_{1}$.
So, dimension of parametric space is 1 in accordance with table in sec.2.

For $m=3$ we have third Kasner hyphersurface, which parametrization
can be obtained from the second one by interchanging $p_{1}$ and $p_{2}$
and by replacement $\theta_{1}\to\theta_{2}$. Cross conditions give else two
equations, additionally to (\ref{eq}). So we have three equtions on
three parameters $\theta_{0},\theta_{1},\theta_{2}$ and then  dimension
of parametric space is 0. Roots of this system of equations have been
founded with the help of special computer program. There is three sets of
solutions:
\begin{equation}
\begin{array}{ccc}
\xi=-1;&\eta=-1;&\mu=-1/2;\\
\\
\xi=-1/2;&\eta=1/2;&\mu=1/2;\\
\\
\xi=1/2;&\eta=-1/2;&\mu=-1;
\end{array}
\end{equation}
where $\mu=\sqrt{2}\cot\theta_{2}$.
It is turn out, that all this solutions reduces by coordinate transformations
to the case $m=1$. In common parametrization all three cases reduces
to one (within redenotions of variables):
\begin{equation}
ds^{2}=dt^{2}-t^{4/3}(dx^{2}+dy^{2})-t^{-2/3}dz^{2}.
\end{equation}
So, new 4-D solution obtained only in the case $m=2$ and can be put
to the following form:
\begin{equation}\label{4gr}
ds^{2}=x^{4q_{0}}dt^{2}-t^{4p_{1}}dx^{2}-t^{4p_{2}}x^{4q_{2}}dy^{2}-t^{4p_{3}}
x^{4q_{3}}dz^{2}
\end{equation}
with
\[
\vec p=\left(
\begin{array}{l}
0\\
\\
1/6-\xi/3\\
\\
1/6(1+\xi)+(1/2\sqrt{3})\sqrt{1-\xi^{2}}\\
\\
1/6(1+\xi)-(1/2\sqrt{3})\sqrt{1-\xi^{2}}
\end{array}
\right);
\]
\[
\vec q=\left(
\begin{array}{l}
1/6-\eta/3\\
\\
0\\
\\
(1+\eta)/6+(1/2\sqrt{3})\sqrt{1-\eta^{2}}\\
\\
(1+\eta)/6-(1/2\sqrt{3})\sqrt{1-\eta^{2}}
\end{array}
\right)
\]
and with equation (\ref{eq}) on parameters $\xi$ and $\eta$.

\section{Conclusion}

In conclusion we make some remarks:

1. The signature of multidimensional interval (\ref{sol}) has no
significance.

2. There are some intersections of obtained solution with one obtained
earlier \cite{sym}.

3. 4-D solution (\ref{4gr}), that have been discussed in last section belong to
a
class of 4-D metric, that admitts abelian $G(2)$ group of isometry.
Moreover its Killing vectors $\partial_{y}, \partial_{z}$ are normal to
hyphersurface.There are some general theorem relating to this class of metric
\cite{sym}. This metric can be related to the static acsially-symmetric
class. One particular solution that corresponds to $\xi=1/2, \eta=-1/2$
belong to the class $BIII$ in classification of vacuum degenerated (type D)
static gravitational fields, proposed by Ehlers and Kundt (see ref. in
\cite{sym} in \S 16.6.2).

4. The question about physical application of obtained solution arises.
In accordance with method given in \cite{kkr}  this solution provided
by the extracoordinates cylindrisity could describe static or evolving
anisotropic mixture of several perfect fluid.


\end{document}